\documentclass[10pt,a4paper]{article}
\usepackage{graphicx}
\usepackage[utf8]{inputenc}     
\usepackage[T1]{fontenc}   
\usepackage{mathptmx}      
\usepackage[left=35mm, right=35mm, top=15mm, bottom=20mm, noheadfoot]{geometry}

\usepackage{listings, color}
\usepackage{cite}
\usepackage{url}
\usepackage{amsfonts}
\begin{document}
\thispagestyle{empty}

\title{\textbf{Softmax Optimizations for Intel$^{\tiny{\textregistered}}$ Xeon$^{\tiny{\textregistered}}$ Processor-based Platforms}}
\author{
Jacek Czaja \\ jacek.czaja@intel.com \\ \textsf{Intel Corporation}
\and
Michal Gallus \\ michal.gallus@intel.com \\ \textsf{Intel Corporation} 
\and
Tomasz Patejko \\ tomasz.patejko@intel.com \\ \textsf{Intel Corporation}
\and
Jian Tang \\ tangjian03@baidu.com \\ \textsf{Baidu}
}

\date{} 
\maketitle\thispagestyle{empty} 
\definecolor{mygray}{rgb}{0.8,0.8,0.8}
\begin{abstract}
This article presents our methodology of optimization and its results applied to softmax\cite{softmax} function. Scope of this work\footnote{See section: Notices and Disclaimers for links to our implementations} includes: algorithmic improvements (reduce general implementation so it is tailored for inference), profiling (identifing most time consuming fragments of code), using efficient computational libraries (Intel$^{\tiny{\textregistered}}$ MKL and Intel$^{\tiny{\textregistered}}$ MKL-DNN) as well as improving vectorization (analysis of applicability of \textsf{OpenMP} and manually crafted assembly code for vectorization improvement). By presenting this methodology, we hope to increase an interest in Deep Learning optimizations for CPUs.

\end{abstract}
\section{Introduction}
Softmax is a function used in classification problems in machine learining, and it has been widely used in deep learning for implementing image classification models, such as AlexNet\cite{alexnet}, GoogleNet\cite{googlenet}, or ResNet\cite{resnet}, where its execution time is small compared to convolution functions. However, natural language processing (NLP) models published recently use softmax more intensively\cite{dam}, and its high computation cost triggered research towards equivalent but computationally cheaper methods, such as hierarchical softmax\cite{hierarchical}.

This article presents optimizations and performance improvements of the softmax operation for x86-64 architectures (in particular Intel$^{\tiny{\textregistered}}$ Xeon$^{\tiny{\textregistered}}$ processors). We focused on inference with deep attention matching (DAM) model, and for our experiments we used Baidu's PaddlePaddle* deep learning platform\cite{paddlepaddle}. We limited our efforts to single-thread execution as the usual optimization process starts with exploiting all the capabilities of a single core. Multithread performance improvements are not the topic of the paper and were explored in\cite{intelcaffe}. We believe that the optimization process presented here could be transfered to other deep learning frameworks such as Tensorflow or PyTorch.

\subsection{Softmax theory}
Softmax function is an extension of logistic regression to work with multiple classification categories.
\begin{equation}
softmax(z_j) = \frac{e^{z_j}}{\sum_{i}^{N} e^{z_i}}
\end{equation}

It is often used as a tool to normalize data. Softmax calculates a probability vector $\mathbf{z}$ for classification tasks where values $z_j$ of vector $\mathbf{z}$ are probability distribution over a set of categories.

\subsection{Softmax as implemented in PaddlePaddle}
Our starting point was PaddlePaddle's softmax implementation using the popular Eigen\cite{eigen} computation library to implement its deep learning operators. Mentioned implementation is presented at Figure \ref{code1}:
\newcommand{\Hilight}{\makebox[0pt][l]{\color{mygray}\rule[-4pt]{0.83\linewidth}{12pt}}}
\lstset{numbers=left}
\begin{figure}[ht]
\begin{lstlisting}[language=C++,basicstyle=\tiny,escapechar=\%]
template <typename DeviceContext, typename T>
void SoftmaxFunctor<DeviceContext, T>::operator()(const DeviceContext& context,
                                                  const framework::Tensor* X,
                                                  framework::Tensor* Y) {
  auto logits = EigenMatrix<T>::From(*X);
  auto softmax = EigenMatrix<T>::From(*Y);

  const int kBatchDim = 0;
  const int kClassDim = 1;

  const int batch_size = logits.dimension(kBatchDim);
  const int num_classes = logits.dimension(kClassDim);

  Eigen::DSizes<int, 1> along_class(kClassDim);
  Eigen::DSizes<int, 2> batch_by_one(batch_size, 1);
  Eigen::DSizes<int, 2> one_by_class(1, num_classes);

  auto shifted_logits = (logits -
                         logits.maximum(along_class)
                             .eval()
                             .reshape(batch_by_one)
                             .broadcast(one_by_class))
                             .unaryExpr(ValueClip<T>());

  softmax.device(*context.eigen_device()) = shifted_logits.exp();
  softmax.device(*context.eigen_device()) = (softmax *
                                             softmax.sum(along_class)
                                                 .inverse()
                                                 .eval()
                                                 .reshape(batch_by_one)
                                                 .broadcast(one_by_class));
}
\end{lstlisting}
\caption{Reference implementation (introduced in PaddlePaddle PR\#14337)}
\label{code1}
\end{figure}

PaddlePaddle does offer a functionality to check the time of execution of operators that were executed. The target CPU used for our work to get performance results of softmax execution was Intel$^{\tiny{\textregistered}}$ Xeon$^{\tiny{\textregistered}}$ Platinum 8180 Processor. We referred to PaddlePaddle profiling\footnote{For details of our experiments that included profiling please refer to appendix: \textsf{Notices and Disclaimers}} to get the performance status of both softmax and overall DAM model, while optimizing softmax. The exemplary profiler report from PaddlePaddle for DAM model execution is presented at figure \ref{prof1} :

\begin{figure}[ht]
\begin{verbatim}
------------------------->     Profiling Report     <-----------------------

Place: CPU
Time unit: ms
Sorted by total time in descending order in the same thread

Event                Calls    Total     Min.      Max.      Ave.       Ratio.      
thread0::layer_norm  316000   68958     0.21599   18.1111   0.218222   0.396    
thread0::softmax     158000   32188.1   0.193633  0.882732  0.203722   0.185    
thread0::stack       19000    19002.9   0.926812  2.51014   1.00015    0.109    
thread0::conv3d      2000     16806.6   1.25346   19.4991   8.40328    0.096   
thread0::mul         317000   13812.1   0.009354  69.8051   0.0435712  0.079   
...
\end{verbatim}
\caption{Exemplary PaddlePaddle's DAM profiling report}
\label{prof1}
\end{figure}

\section{Optimization process}
\subsection{Algorithmic modifications}
By inspecting the implementation code we can see that there is a functionality of ValueClip. When ValueClip is used, softmax does not produce zero values by assiging very small floating point constant eg. $10^{-60}$ to a variable that holds zero. This is needed in a situation when there is some logarithmic operation following softmax eg. cross-entropy loss. A logarithm of 0 is -inf which will later on produce NAN in a training. However as we are optimizing(speeding up) inference adding this minimal threshold is not needed.  After removing mentioned functionality \textbf{execution time is reduced by 5\%}

\subsection{Profiling}
Once we analyzed softmax implementation and removed unnecessary elements, we can do profile operations in the softmax operator to find operations that are the most time consuming. Hotspots are the operations on which we focus our attention because their successful optimization can potentially give the highest performance gains.
\lstset{numbers=left}
\begin{figure}[ht]
\begin{lstlisting}[language=C++,basicstyle=\tiny,escapechar=\%]
  %\Hilight%#include <x86intrin.h>
....
template <typename DeviceContext, typename T>
void SoftmaxFunctor<DeviceContext, T>::operator()(const DeviceContext& context,
                                                  const framework::Tensor* X,
                                                  framework::Tensor* Y) {
  auto logits = EigenMatrix<T>::From(*X);
  auto softmax = EigenMatrix<T>::From(*Y);

  const int kBatchDim = 0;
  const int kClassDim = 1;

  const int batch_size = logits.dimension(kBatchDim);
  const int num_classes = logits.dimension(kClassDim);

  Eigen::DSizes<int, 1> along_class(kClassDim);
  Eigen::DSizes<int, 2> batch_by_one(batch_size, 1);
  Eigen::DSizes<int, 2> one_by_class(1, num_classes);

  %\Hilight% unsigned long long t0 = __rdtsc();
  auto shifted_logits = (logits -
                         logits.maximum(along_class)
                             .eval()
                             .reshape(batch_by_one)
                             .broadcast(one_by_class));

  %\Hilight% unsigned long long t1 = __rdtsc();
  softmax.device(*context.eigen_device()) = shifted_logits.exp();
  %\Hilight% unsigned long long t2 = __rdtsc();
  softmax.device(*context.eigen_device()) = (softmax *
                                             softmax.sum(along_class)
                                                 .inverse()
                                                 .eval()
                                                 .reshape(batch_by_one)
                                                 .broadcast(one_by_class));
  %\Hilight% unsigned long long t3 = __rdtsc();

  %\Hilight% std::cout << "softmax computing time is :" << (t3-t0) << std::endl;
  %\Hilight% std::cout << "shifted_logits computing is :" << (t1-t0)/((float)(t3-t0)) << " of softmax time" << std::endl;
  %\Hilight% std::cout << "exp computing is :" << (t2-t1)/((float)(t3-t0)) << " of softmax time" << std::endl;
  %\Hilight% std::cout << "sum&div computing is :" << (t3-t2)/((float)(t3-t0)) << " of softmax time" << std::endl;
}
\end{lstlisting}
\caption{Reference timed implementation}
\label{code2}
\end{figure}

We used timestamp counter (TSC), which is very precise time measuring device for CPUs. Instruction \textsf{\_\_rdtsc} returns the current value of this CPU clock. Idea is that we measure the time of entire execution of operator as well as selected parts of it, so we know which part takes the most time relatively. When we begin modification/optimization the absolute value will also be useful to see if we are making progress. One note is that profiling is done on more than one executions of function for more reliable results.
The rough idea of profiling is presented in listing \ref{code2}. After finished execution, profiling told us that: over 50\% of time of softmax execution is spent in exp part of function and sum\&div took around ~30\%. Hence optimization of $e^x$ followed by summing and elementwise division would be our targets. 

\subsection{Performance improvements with Intel$^{\tiny{\textregistered}}$ Math Kernel Library (Intel$^{\tiny{\textregistered}}$ MKL)}
To spare developers the effort of low level optimizations for the most common mathematical algorithms, a number of libraries have been created that provide optimized implementations of such operations: OpenBLAS\cite{openblas}, Eigen\cite{eigen} and Intel$^{\tiny{\textregistered}}$ MKL. PaddlePaddle baseline code does use Eigen which is fast and elegant library, but we use Intel$^{\tiny{\textregistered}}$ MKL as it provides implementations optimized for x86\_64 architectures (in particular Intel$^{\tiny{\textregistered}}$ Xeon$^{\tiny{\textregistered}}$ processors). We replaced exponential computations and elementwise division with BLAS functions provided by Intel$^{\tiny{\textregistered}}$ MKL and the remaining Eigen code was replaced with a manually hand crafted implementation[see figure \ref{code3}]. \textbf{Performance improvement was around 2X.}

\lstset{numbers=left}
\begin{figure}[ht]
\begin{lstlisting}[language=C++,basicstyle=\tiny,escapechar=\%]
template <typename DeviceContext>
class SoftmaxFunctor<DeviceContext, float, true> {
  void operator()(const DeviceContext& context, const framework::Tensor* X,
                  framework::Tensor* Y) {

    auto in_dims = X->dims();
    auto out_dims = Y->dims();
    const float* in_data = X->data<float>();
    float* out_data = Y->data<float>();
    const int kBatchDim = 0;
    const int kClassDim = 1;
    // 2D data. Batch x C
    const int batch_size = in_dims[kBatchDim];
    const int num_classes = in_dims[kClassDim];
    for (int n=0; n < batch_size; ++n) {
      float max = in_data[n*num_classes];
      for (int c=1; c < num_classes; ++c) {
        max = in_data[n*num_classes + c] > max ? in_data[n*num_classes+c] : max;
      }
      for (int c=0; c < num_classes; ++c) {
        out_data[n*num_classes+c] = in_data[n*num_classes+c] - max;
      }
    }
    vsExp(num_classes*batch_size, out_data, out_data);

    for (int n=0; n < batch_size; ++n) {
      float sum = out_data[n*num_classes];
      for (int c=1; c < num_classes; ++c) {
        sum += out_data[n*num_classes + c];
      }
      cblas_sscal(num_classes, 1.0f/sum, &out_data[n*num_classes], 1);
    }
  }
};
}
\end{lstlisting}
\caption{MKL based implementation (introduced in PaddlePaddle PR\#14437)}
\label{code3}
\end{figure}

\subsection{Auto-vectorization with OpenMP}
Our next step was to improve the code that was not too be replaced with Intel$^{\tiny{\textregistered}}$ MKL. We optimized the following operations:
\begin{itemize}
\item Subtracting value from all elements of vector (elementwise subtraction)
\item Summing up vector elements
\item Finding the maximal value within elements of vector
\end{itemize}

We took advantage of \textsf{OpenMP simd reduction} clause\cite{openmp} that was introduced in OpenMP 4.0 and is available in ICC and GCC (from version 4.9).
OpenMP \textsf{simd} can be seen as a way to provide additional details on a implementation and reduction mechanism so a compiler can more effectively vectorize the code.  
More detailed information on OpenMP vectorization can be found\cite{openmp-simd}.

\subsubsection{Elementwise subtraction}
We inspected generated assembly of the already mentioned three operations\footnote{See Appendix for description of methodology used} and found that elementwise subtraction is already vectorized. So no further hints to the compiler were needed.

\subsubsection{Summing Up Elements}
We found that OpenMP \textsf{simd} by itself (hints to loops vectorization) did not provide much of a performance boost. It may result in code size reduction as the compiler did not have to generate multiple implementations of code, when some hints were provided. On the other hand \textsf{OpenMP simd} followed by \textsf{reduction} clause decreased execution time significantly.

Figure \ref{code4} presents original assembly of summing procedure, as generated by the compiler. It can be seem that although the Intel$^{\tiny{\textregistered}}$ AVX instruction set is used (eg.
\textsf{vaddss} is used) it does not operate on 128/256 bit words, it just adds sequentially 32-bit words.

\lstset{numbers=left}
\begin{figure}[ht]
\begin{lstlisting}[language=C++,basicstyle=\tiny,escapechar=\%]
	BEGIN SEQUENCE SUM! <---
# 0 "" 2
#NO_APP
	vxorps	xmm0, xmm0, xmm0
	lea	eax, [rdx-1]
	test	edx, edx
	lea	rax, [rsi+4+rax*4]
	vmovss	DWORD PTR [rdi], xmm0
	jle	.L3
	.p2align 4,,10
	.p2align 3
.L4:
	vaddss	xmm0, xmm0, DWORD PTR [rsi]
	add	rsi, 4
	cmp	rax, rsi
	vmovss	DWORD PTR [rdi], xmm0
	jne	.L4
.L3:
#APP
# 36 "/home/jacekc/test-openmp/main.cpp" 1
	END SEQUENCE SUM! <---
\end{lstlisting}
\caption{Assembly code of reduction(summing up) of vector}
\label{code4}
\end{figure}

The highlighted line of code in figure \ref{code5} contains modification we introduced. By marking the loop in the line below with pragma \textsf{omp simd reduction}, we gave a hint to the compiler that reduction on variable \texttt{sum} can be safely vectorized, and each partial sum can be computed in parallel using Intel$^{\tiny{\textregistered}}$ AVX instructions. We inspected the generated assembly (see Figure \ref{code7}) to check that introduced modification brought expected vectorization.\footnote{Extract of generated assembly shows packed vector AVX instructions that implement sum reduction, that perform computations on 128 bit memory chunks eg. \textsf{vaddps}} 

\lstset{numbers=left}
\begin{figure}[ht]
\begin{lstlisting}[language=C++,basicstyle=\tiny,escapechar=\%]
template <typename DeviceContext>
class SoftmaxFunctor<DeviceContext, float, true> {
  void operator()(const DeviceContext& context, const framework::Tensor* X,
                  framework::Tensor* Y) {

    auto in_dims = X->dims();
    auto out_dims = Y->dims();
    const float* in_data = X->data<float>();
    float* out_data = Y->data<float>();
    const int kBatchDim = 0;
    const int kClassDim = 1;
    // 2D data. Batch x C
    const int batch_size = in_dims[kBatchDim];
    const int num_classes = in_dims[kClassDim];
    for (int n=0; n < batch_size; ++n) {
      float max = in_data[n*num_classes];
      for (int c=1; c < num_classes; ++c) {
        max = in_data[n*num_classes + c] > max ? in_data[n*num_classes+c] : max;
      }
      for (int c=0; c < num_classes; ++c) {
        out_data[n*num_classes+c] = in_data[n*num_classes+c] - max;
      }
    }
    vsExp(num_classes*batch_size, out_data, out_data);

    for (int n=0; n < batch_size; ++n) {
      float sum = out_data[n*num_classes];
%\Hilight%      #pragma omp simd reduction(+ : sum)
      for (int c=1; c < num_classes; ++c) {
        sum += out_data[n*num_classes + c];
      }
      cblas_sscal(num_classes, 1.0f/sum, &out_data[n*num_classes], 1);
    }
  }
};
}
\end{lstlisting}
\caption{MKL and openmp simd  based implementation}
\label{code5}
\end{figure}
This optimization brought an additional 5\% reduction in time execution.

Although there was a performance improvement, we did not use it in PaddlePaddle. Summing $e^{z_i}$ is an operation that sums positive values, and Intel$^{\tiny{\textregistered}}$ MKL already provides such operation, \texttt{cblas\_sasum}, that sums absolute values of elements. The advantage of using MKL's sasum is that OpenMP pragmas support is present in the recent generation of compilers, but in production environments some old compilers like MSVC and GCC 4.8 are still used so when using Intel$^{\tiny{\textregistered}}$ MKL we speed up also those older configurations. Fortunately our OpenMP vectorization effort was not discarded, as we upstreamed it into the Intel$^{\tiny{\textregistered}}$ MKL-DNN project\footnote{MKL-DNN can be build either with Intel MKL as well as without} softmax implementation.

\lstset{numbers=none}
\begin{figure}[ht]
\begin{lstlisting}[language=C++,basicstyle=\tiny,escapechar=\%]
..............
	vmovaps	ymm0, YMMWORD PTR [rbp-48]
	xor	r8d, r8d
.L11:
	mov	r9, r8
	add	r8, 1
	sal	r9, 5
	cmp	ecx, r8d
	vaddps	ymm0, ymm0, YMMWORD PTR [rsi+r9]
	ja	.L11
	cmp	eax, edx
	vmovaps	YMMWORD PTR [rbp-48], ymm0
.............
\end{lstlisting}
\caption{Fragment of Assembly code of vectorized reduction(summing up) of vector}
\label{code6}
\end{figure}

\subsubsection{Finding maximal value in an array of elements}
We applied \textsf{openmp simd reduction(max:)} for searching maximal value in softmax operator, but despite asking the compiler for max reduction, the generated code was not vectorized. Hence we implemented max value search directly using assembly language (see \ref{manual}).

\subsection{Vectorization with SIMD instructions} \label{manual}
As we have showed here, compiler auto-vectorization capabilities, when used carefully, can bring visible performance improvements. However, that is not always the case because code generated by the compiler's auto-vectorizer can turn out to be suboptimal, and the only option is to implement an algorithm manually with SIMD instructions (AVX, AVX2 and AVX512). So performance critical functionality may benefit significantly when implemented manually in assembly, in particular when vector instructions need to be used.

Assembly implementation of max value search was implemented with a help of Xbyak\cite{xbyak} project. Xbyak is a JIT assembler that generates assembly code at runtime. JIT functionality  suits deep learning use cases very well, as declared models (description of neural network) are usually not modified during their execution (inference, training). Hence, we can generate assembly after the model was defined and we can have assembly code suited for a neural network model. In particular we can have different assembly code for different batch sizes.

The manually crafted assembly code for finding maximal value in an array is presented at figure \ref{code9}. Due to the introduction of a manually crafted assembler, the max finding function is around three times faster than our reference code. As the percentage of time spent executing max function is small compared to the computation of exponential function $e^x$, its performance impact on softmax is small. Softmax after implementing max finding value in Xbyak is on average 3\% faster. It may seem small, but for data centers that are constantly executing deep learning workloads, even 3\% improvement can account for a significant savings in energy and time.

Portability and maintainability problems are the main disadvantages of using assembly language for performance optimizations. Assembly code is not portable among different architectures, and it is more difficult to maintain than the implementations written in higher-level languages. In general, if possible, we recommend using existing softmax implementations like those provided by MKL-DNN, and implementing critical operations with assembly language only when they are not available in Intel$^{\tiny{\textregistered}}$ MKL-DNN or other fast computational libraries.

\subsection{Limits of optimizations}
When working on optimizing the code we would like to know if there is any room for improvement of its execution time. We need a measure of how close is actual performance of softmax to platform maximal capabilities. There are two limitations to improving performance on given hardware platform eg.
\begin{itemize}
\item memory bound limit
\item computation bound limit
\end{itemize} 

Those two limitations and kernel's operational intensity are foundation for Roofline model\cite{roofline}, which is often used for estimation if further performance optimizations are possible. The application of Roofline model is out of scope of this document and was not used during the work discussed here. When working on softmax optimization in the context of a DAM model, we experimented by replacing softmax computation with memory copying routine eg. memcpy. Memcpy is usually well optimized (often written manually using vector instructions) so the speed of execution of memcpy is limited by memory throughput. Softmax takes some input buffer and writes its result to output buffer, but both buffers are of the same size hence memcpy can be used to replace softmax computation. The comparison of both execution times (actual softmax implementation and memcpy) can give us an idea of whether it is worth to invest more time into the optimization of softmax implementation. If the softmax execution time is close to memcpy then it is likely that the algorithm is bound by maximal memory throughput and we won't get better performance in a given execution environment. We initially verified (using memcpy) that baseline (not fully vectorized) implementation is far from being memory bound (see Figure \ref{performance1}) and based on that result we concluded that performance can be increased by better utilisation of computing resources of processor eg. Introducing effective Intel$^{\tiny{\textregistered}}$ MKL implementations and more effective vectorization.

\section{Performance Evaluation} \label{performance_evaluation}

\begin{figure}[ht]
\input{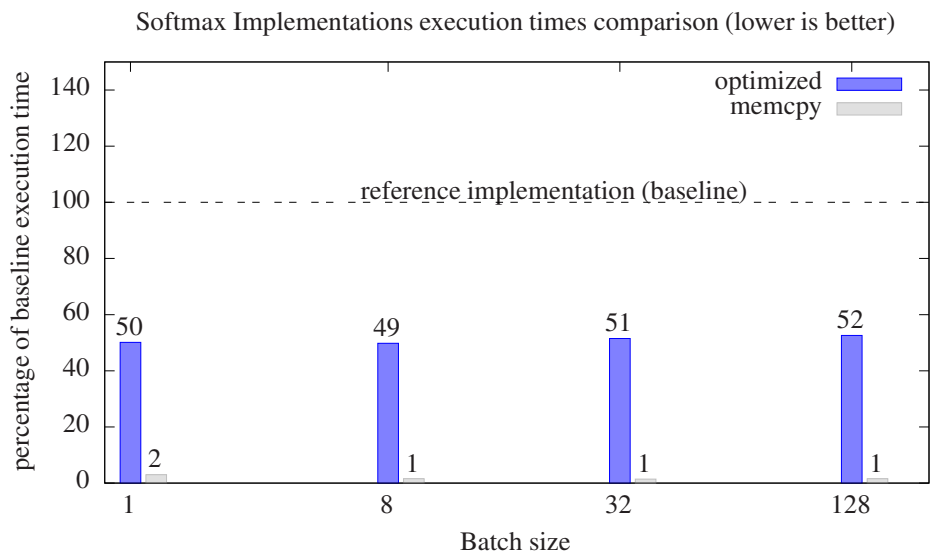}
\caption{Softmax implementations performance comparison}
\label{performance1}
\end{figure}

\begin{figure}[ht]
\input{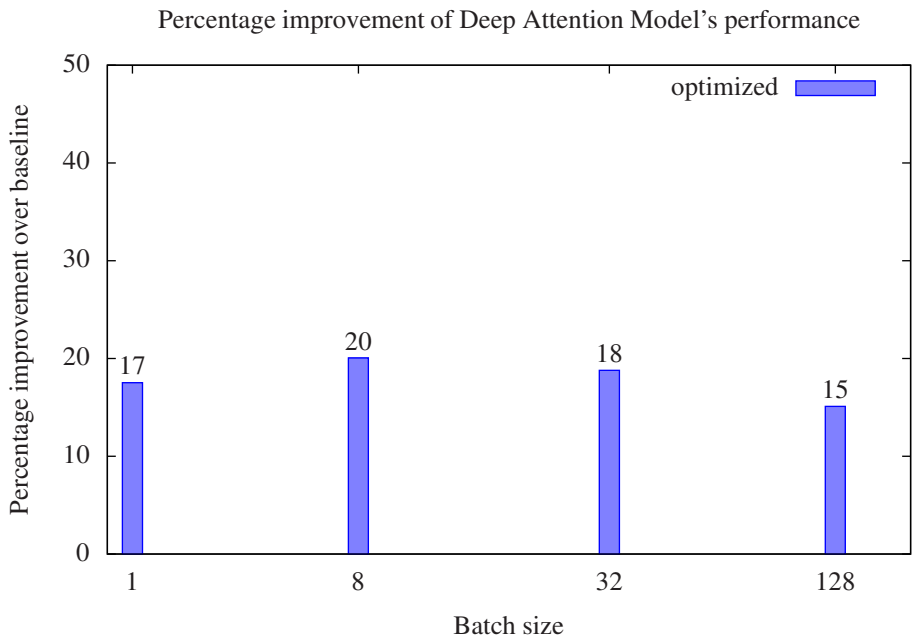}
\caption{DAM models overall performance comparison}
\label{performance2}
\end{figure}

Figure \ref{performance1} shows that the softmax execution in DAM model is 2X faster than the original implementation. This optimization impacts performance of the entire DAM model and improves it by over 15\% (figure \ref{performance2}).

\section{Conclusion and Further Work}
We presented the methods we used to optimize softmax as well as demonstrated the performance gain as a result of this methodology. 

From profiling information, we observed that exponential functions execution take up significant amount of time. This could be further improved by using cheaper in execution approximation functions which can provide a performance boost in exchange for little computational inaccuracy. Another idea could be applying a roofline model to softmax implementations to get an estimation of how much more performance could potentially be improved. Also, knowing that optimized implementation is far away from memory throughput limitation, it would be beneficial to use the Intel$^{\tiny{\textregistered}}$ AVX512 instruction set to manually implement the entire Softmax operator. As softmax is a popular deep learning primitive, we upstreamed\footnote{Sum reduction using both OpenMP vectorization MKL was merged to MKL-DNN codebase} our optimizations to the Intel$^{\tiny{\textregistered}}$ MKL-DNN library.

\section{Acknowledgments}
The authors would like to express our gratitude to Krzysztof Badziak, Mateusz Ozga and Evarist Fomenko from Intel Corporation for their advice on optimizations, and to Luo Tao from Baidu Corporation for her helpful suggestions.

\bibliography{articles}{}
\bibliographystyle{plain}

\section*{Appendix: Inspecting generated assembly code}

For the purpose of a work presented here generated assembly code was inspected using compiler switches (relevant to GCC):
\begin{verbatim}
-S -masm=intel
\end{verbatim}

As well as injecting markers into code to easily locate section of code that we are interested in. For example
following C++ code (Figure \ref{code7}) , after applying mentioned compiler switches results in generated assembly as presented on Figure \ref{code8}.

\lstset{numbers=left}
\begin{figure}[ht]
\begin{lstlisting}[language=C++,basicstyle=\tiny,escapechar=\%]
#     ifdef GENERATE_ASSEMBLY
      asm volatile ("BEGIN SIMD SOFTMAX SUM! <---");
#     endif
      float* tmpptr = &out_data[n*num_classes];
      #pragma omp simd reduction(+: result) aligned(tmpptr)
      for (int c=0; c < num_classes; ++c) {
        result += tmpptr[c];
      }
      entities[n] = result; 
#     ifdef GENERATE_ASSEMBLY
      asm volatile ("END SIMD SOFTMAX SUM! <---");
#     endif
\end{lstlisting}
\caption{Assembly code of vectorized reduction(summing up) of vector}
\label{code7}
\end{figure}

\lstset{numbers=left}
\begin{figure}[ht]
\begin{lstlisting}[language=C++,basicstyle=\tiny,escapechar=\%]
	BEGIN SIMD SUM! <---
# 0 "" 2
#NO_APP
	test	edx, edx
	mov	DWORD PTR [rdi], 0x00000000
	mov	QWORD PTR [rbp-48], 0
	mov	QWORD PTR [rbp-40], 0
	mov	QWORD PTR [rbp-32], 0
	mov	QWORD PTR [rbp-24], 0
	jle	.L9
	lea	ecx, [rdx-8]
	shr	ecx, 3
	add	ecx, 1
	cmp	edx, 7
	lea	eax, [0+rcx*8]
	jle	.L15
	vmovaps	ymm0, YMMWORD PTR [rbp-48]
	xor	r8d, r8d
.L11:
	mov	r9, r8
	add	r8, 1
	sal	r9, 5
	cmp	ecx, r8d
	vaddps	ymm0, ymm0, YMMWORD PTR [rsi+r9]
	ja	.L11
	cmp	eax, edx
	vmovaps	YMMWORD PTR [rbp-48], ymm0
	je	.L21
	vzeroupper
.L10:
	movsx	rcx, eax
	vmovss	xmm0, DWORD PTR [rsi+rcx*4]
	lea	ecx, [rax+1]
	vaddss	xmm0, xmm0, DWORD PTR [rbp-48]
	cmp	edx, ecx
	vmovss	DWORD PTR [rbp-48], xmm0
	jle	.L9
	movsx	rcx, ecx
	vaddss	xmm0, xmm0, DWORD PTR [rsi+rcx*4]
	lea	ecx, [rax+2]
	cmp	edx, ecx
	vmovss	DWORD PTR [rbp-48], xmm0
	jle	.L9
	movsx	rcx, ecx
	vaddss	xmm0, xmm0, DWORD PTR [rsi+rcx*4]
	lea	ecx, [rax+3]
	cmp	edx, ecx
	vmovss	DWORD PTR [rbp-48], xmm0
	jle	.L9
	movsx	rcx, ecx
	vaddss	xmm0, xmm0, DWORD PTR [rsi+rcx*4]
	lea	ecx, [rax+4]
	cmp	edx, ecx
	vmovss	DWORD PTR [rbp-48], xmm0
	jle	.L9
	movsx	rcx, ecx
	vaddss	xmm0, xmm0, DWORD PTR [rsi+rcx*4]
	lea	ecx, [rax+5]
	cmp	edx, ecx
	vmovss	DWORD PTR [rbp-48], xmm0
	jle	.L9
	movsx	rcx, ecx
	add	eax, 6
	vaddss	xmm0, xmm0, DWORD PTR [rsi+rcx*4]
	cmp	edx, eax
	vmovss	DWORD PTR [rbp-48], xmm0
	jle	.L9
	cdqe
	vaddss	xmm0, xmm0, DWORD PTR [rsi+rax*4]
	vmovss	DWORD PTR [rbp-48], xmm0
.L9:
	vxorps	xmm0, xmm0, xmm0
	vaddss	xmm0, xmm0, DWORD PTR [rbp-48]
	vaddss	xmm0, xmm0, DWORD PTR [rbp-44]
	vaddss	xmm0, xmm0, DWORD PTR [rbp-40]
	vaddss	xmm0, xmm0, DWORD PTR [rbp-36]
	vaddss	xmm0, xmm0, DWORD PTR [rbp-32]
	vaddss	xmm0, xmm0, DWORD PTR [rbp-28]
	vaddss	xmm0, xmm0, DWORD PTR [rbp-24]
	vaddss	xmm0, xmm0, DWORD PTR [rbp-20]
	vmovss	DWORD PTR [rdi], xmm0
#APP
# 52 "/home/jacekc/test-openmp/main.cpp" 1
	END SIMD SUM! <---
\end{lstlisting}
\caption{Assembly code of vectorized reduction(summing up) of vector}
\label{code8}
\end{figure}

\lstset{numbers=left}
\begin{figure}[ht]
\begin{lstlisting}[language=C++,basicstyle=\tiny,escapechar=\%]
struct maxUFunc : public Xbyak::CodeGenerator {
    maxUFunc()
{
#if defined(__x86_64__)
// calling convention RDI, RSI, RDX, RCX, R8, R9
// XMM0-7 (ints are passed that way)
//      RDI - Reference to Result
//      RSI - PTR to Array
//      RDX - Num classes 

// Regsters that need to be preserved: RBX,RBP, R12-R15

  Xbyak::util::Cpu current_cpu;
  if(current_cpu.has(Xbyak::util::Cpu::tAVX2)) {
    printf("AVX2 supported!\n");
  } else {
    printf("AVX2 not detected!\n");
  }

  mov (rcx,rdx);	
  push(rbx);
  shr (rcx,3);  // Divide by 8 (eight floats)
  shl (rdx,2);  // num of Output elements * size of float (4)
  shl (rcx,5);  // Trunc to 32 bytes 


	// Compute partial maximums
  vpbroadcastd(ymm0,ptr [rsi]);
  xor(rax,rax);				// Move offset for next 8 floating point values
  L("for_i");
    cmp(rax,rcx);
    jz("tail");
    vmovups(ymm1,ptr [rsi + rax]);  // A
		add(rax,32);				// Move offset for next 8 floating point values
		vmaxps(ymm0,ymm0,ymm1);
    jmp("for_i");
  // Tail execution
  L("tail");
    sub(rdx,rcx);
    cmp(rdx,16);  
    jb("seq");
    vmovups(xmm2,ptr [rsi + rax]);  // A
		add(rax,16);				// Move offset for next 4 floating point values
    sub(rdx,16);
		vperm2f128(ymm2,ymm2,ymm2,0);
		vmaxps(ymm0,ymm0,ymm2);  //partial maxes in ymm0
  L("seq");
	  cmp(rdx,0);
    jz("done");	
		vpbroadcastd(ymm2,ptr [rsi + rax]);
		vmaxps(ymm0,ymm0,ymm2);  //partial maxes in ymm0
    sub(rdx,4);
    add(rax,4);
    jmp("seq");
  L("done");
  // Get within shortlisted buffer maximum
	vperm2f128(ymm1,ymm0,ymm0,1);
  vmaxps(ymm0,ymm0,ymm1);  //partial maxes in ymm0
  vpermilps(xmm1,xmm0,0x1B);
  vmaxps(ymm0,ymm0,ymm1);  //partial maxes in ymm0
  vpermilps(xmm1,xmm0,1);
  vmaxps(ymm0,ymm0,ymm1);  //ymm0[0:31] contains global maximum
  vmovss(ptr[rdi],xmm0); // Result <-Max(X[.])
  pop(rbx);

  printf("Generating Max Value code\n");
#else
        printf("32bit not supported\n");
#endif
  ret();
}
};
\end{lstlisting}
\caption{JIT Assembly code of maximal value finding in an array}
\label{code9}
\end{figure}

\section*{Appendix: Notices and Disclaimers}
Intel$^{\tiny{\textregistered}}$ technologies features and benefits depend on system configuration and may require enabled hardware, software or service activation. Performance varies depending on system configuration.\\\\

Most of our work was upstreamed into PaddlePaddle and MKL-DNN projects. And can be accessed respectively at:\\\\
\url{https://github.com/PaddlePaddle/Paddle}\footnote{All quoted Pull Requests within this article, are related to this repository }\\\\
and \\\\
\url{https://github.com/intel/mkl-dnn}\\\\
However, optimizations of softmax using direct implementation in assembly language (section \ref{manual}) at the moment of writing this article are not
part of PaddlePaddle and MKL-DNN repositories. Hence for taking performance measures We created an integration branch located at:\\\\
\url{https://github.com/tpatejko/Paddle/commits/tpatejko/jit-max-in-softmax}\\\\
The Experiments were executed\footnote{using commit \textsf{28bba75d9108026f236c312813caf5ba72a6aabe} from integration branch} using following commands:
\begin{verbatim}
OMP_NUM_THREADS=1 ./paddle/fluid/inference/tests/api/test_analyzer_dam \
   --infer_model=third_party/inference_demo/dam/model/ \
   --infer_data=third_party/inference_demo/dam/data.txt \
   --gtest_filter=Analyzer_dam.profile --batch_size=1 \
   --test_all_data=true --num_threads=1 --use_analysis=false --profile
echo "===> Batch 8"
OMP_NUM_THREADS=1 ./paddle/fluid/inference/tests/api/test_analyzer_dam \
   --infer_model=third_party/inference_demo/dam/model/ \
   --infer_data=third_party/inference_demo/dam/data.txt \
   --gtest_filter=Analyzer_dam.profile --batch_size=8 \
   --test_all_data=true --num_threads=1 --use_analysis=false --profile
echo "===> Batch 32"
OMP_NUM_THREADS=1 ./paddle/fluid/inference/tests/api/test_analyzer_dam \
   --infer_model=third_party/inference_demo/dam/model/ \
   --infer_data=third_party/inference_demo/dam/data.txt \
   --gtest_filter=Analyzer_dam.profile --batch_size=32 \
   --test_all_data=true --num_threads=1 --use_analysis=false --profile
echo "===> Batch 128"
OMP_NUM_THREADS=1 ./paddle/fluid/inference/tests/api/test_analyzer_dam \
   --infer_model=third_party/inference_demo/dam/model/ \
   --infer_data=third_party/inference_demo/dam/data.txt \
   --gtest_filter=Analyzer_dam.profile --batch_size=128 \
   --test_all_data=true --num_threads=1 --use_analysis=false --profile
\end{verbatim}

\end{document}